\documentclass[twoside]{article}

\usepackage[accepted]{aistats2020}

\usepackage[round]{natbib}

\usepackage[title]{appendix}
\usepackage[utf8]{inputenc} %
\usepackage[T1]{fontenc}    %
\usepackage{lmodern}
\usepackage{hyperref}       %
\usepackage{url}            %
\usepackage{booktabs}       %
\usepackage{amsfonts}       %
\usepackage{nicefrac}       %
\usepackage{microtype}      %
\usepackage{amsmath}
\usepackage{amssymb}
\usepackage{graphicx}
\usepackage{algorithm}
\usepackage{algorithmicx}
\usepackage{algpseudocode}
\usepackage[nameinlink]{cleveref}
\creflabelformat{equation}{#2\textup{#1}#3}  %

\newcommand{\bbE}{\mathbb{E}}
\newcommand{\bbR}{\mathbb{R}}
\newcommand{\unif}{\mathrm{Unif}}
\newcommand{\calZ}{\mathcal{Z}}
\newcommand{\calN}{\mathcal{N}}
\newcommand{\Bern}{\mathrm{Bern}}
\newcommand{\Gam}{\mathrm{Gam}}
\newcommand{\Var}{\mathop{\textrm{Var}}}
\newcommand{\Cov}{\mathop{\textrm{Cov}}}
\crefname{appsection}{Appendix}{Appendices}
\algrenewcommand\algorithmicindent{0.5em}
\algnewcommand{\LineComment}[1]{\State \(\triangleright\) #1}

\begin{document}

\twocolumn[

\aistatstitle{Hamiltonian Monte Carlo Swindles}

\aistatsauthor{ Dan Piponi \And Matthew D. Hoffman \And Pavel Sountsov }

\aistatsaddress{ Google Research \And Google Research \And Google Research } ]

\begin{abstract}

Hamiltonian Monte Carlo (HMC) is a powerful Markov chain Monte Carlo (MCMC)
algorithm for estimating expectations with respect to continuous
un-normalized probability distributions. MCMC estimators typically have
higher variance than classical Monte Carlo with i.i.d. samples due to
autocorrelations; most MCMC research tries to reduce these
autocorrelations. 
In this work, we explore a complementary approach to variance
reduction based on two classical Monte Carlo ``swindles'': first, running an
auxiliary coupled chain targeting a tractable approximation to the target
distribution, and using the auxiliary samples as control variates; and
second, generating anti-correlated ("antithetic") samples by running two
chains with flipped randomness. Both ideas have been explored previously
in the context of Gibbs samplers and random-walk Metropolis algorithms,
but we argue that they are ripe for adaptation to HMC in light of recent
coupling results from the HMC theory literature.
For many posterior distributions, we find that
these swindles generate effective sample sizes orders of magnitude larger
than plain HMC, as well as being more efficient than analogous swindles for
Metropolis-adjusted Langevin algorithm and random-walk Metropolis.

\end{abstract}

\section{Introduction}

Recent theoretical results demonstrate that one can couple two
Hamiltonian Monte Carlo \citep[HMC; ][]{neal2011mcmc} chains by
feeding them the same random numbers; that is, even if the two chains
are initialized with different states the two chains' dynamics will
eventually become indistinguishable \citep{mangoubi,bou-rabee}. These
results were proven in service of proofs of convergence and mixing
speed for HMC, but have been used by \citet{heng2019unbiased} to
derive unbiased estimators from finite-length HMC chains.

Empirically, we observe that two HMC chains targeting slightly
different stationary distributions still couple \emph{approximately}
when driven by the same randomness. In this paper, we propose methods
that exploit this approximate-coupling phenomenon to reduce the
variance of HMC-based estimators. These methods are based on two
classical Monte Carlo variance-reduction techniques (sometimes called
``swindles''): antithetic sampling and control variates.

In the antithetic scheme, we induce anti-correlation between two
chains by driving the second chain with momentum variables that are
negated versions of the momenta from the first chain. When the chains
anti-couple strongly, averaging them yields estimators with very low
variance. This scheme is similar in spirit to the antithetic-Gibbs
sampling scheme of \citet{frigessi2000antithetic}, but is applicable
to problems where Gibbs may mix slowly or be impractical due to
intractable complete-conditional distributions.

In the control-variate scheme, following \citet{pinto2001improving}
we run an auxiliary chain targeting a
tractable approximation to the true target distribution (i.e., a
distribution for which computing or estimating expectations is cheap),
and couple this auxiliary chain to the original HMC chain.  The
auxiliary chain can be plugged into a standard control-variate
swindle, which yields low-variance estimators to the extent that the
two chains are strongly correlated.

These two schemes can be combined to yield further variance
reductions, and are especially effective when combined with powerful
preconditioning methods such as transport-map MCMC
\citep{parno2014transport} and NeuTra \citep{hoffman2019neutra}, which
not only reduce autocorrelation time but also increase coupling speed.
When we can approximate the target distribution well, our swindles can
reduce variance by an order of magnitude or more, yielding
super-efficient estimators with effective sample sizes significantly
larger than the number of steps taken by the underlying Markov chains.

\section{Hamiltonian Monte Carlo Swindles}
\subsection{Antithetic Variates}
Suppose $X$ is a random variable following some distribution $P(x)$ and we wish to estimate $\bbE_P[f(X)]$.
If we have two (not necessarily independent) random variables $X_1, X_2$ with marginal distribution $P$ then
\begin{align*}
        \Var\left[\frac{f(X_1)+f(X_2)}{2}\right] =& \frac{1}{4} \big( \Var[f(X_1)] + \Var[f(X_2)]\\
        &+2\Cov[f(X_1),f(X_2)]\big)
\end{align*}
When $f(X_1)$ and $f(X_2)$ are negatively correlated the average of these two random variables will have a lower variance than the average of two independent draws from the distribution of $X$.
We will exploit this identity to reduce the variance of the HMC algorithm.

First we consider the usual HMC algorithm to generate samples from the probability distribution with density function $P(x) \triangleq \calZ^{-1}\exp(-U(x))$ on $\bbR^D$ \citep{neal2011mcmc}.
The quantity $\calZ$ is a possibly unknown normalization constant.
Let $\calN(0, I_D)$ be the standard multivariate normal distribution on
$\bbR^d$ with mean $0$ and covariance $I_D$ (the identity matrix in $D$
dimensions).
For position and momentum vectors $q$ and $p$ each in $\bbR^D$ define the Hamiltonian $H(q,p)=\frac{1}{2}||p||^2+U(q)$.
Also suppose that $\Psi_{H,T}(q,p)$ is a symplectic integrator \citep{hairer} that numerically simulates the physical system described by the Hamiltonian $H$ for time $T$, returning $q', p'$.
A popular such method is the leapfrog integrator.
The standard Metropolis-adjusted HMC algorithm is:
\begin{algorithmic}[1]
    \Procedure{HMC}{$X_0, n, T$}
        \For{$i=0$ to $n-1$}
            \LineComment{Sample momentum from normal distribution}
            \State Sample $p_i \sim \calN(0, I_D)$.
            \LineComment{Symplectic integration}
            \State Set $(q_{i+1}, p_{i+1}) = \Psi_{H,T}(X_i, p_i)$.
            \State Sample $b_i \sim \unif([0,1])$.
            \State Set $X_{i + 1}$ = \textsc{MHAdj}($X_i$, $q_{i+1}$, $p_i$, $p_{i+1}$, $b_i$, $H$)
        \EndFor
    \EndProcedure
    \Procedure{MHAdj}{$q_0$, $q_1$, $p_0$, $p_1$, $b$, $H$}
        \If{$b < \min(1,\exp(H(q_0,p_0)-H(q_{1},p_{1})))$}
            \State \Return $q_1$.
                \Else
            \State \Return $q_0$.
        \EndIf
    \EndProcedure
\end{algorithmic}
The output is a sequence of samples $X_1,\ldots,X_n$ that converges to $P(x)$ for large $n$ \citep{neal2011mcmc}.

Next, we consider a pair of coupled HMC chains:
\begin{algorithmic}[1]
    \Procedure{HMC-COUPLED}{$X_0, Y_0, n, T$}
        \For{$i=0$ to $n-1$}
            \State Sample $p_i \sim \calN(0, I_D)$.
            \State Set $p'_i=p_i$.
            \label{momentum}
            \Comment{Shared $p_i$}
            \State Set $(q_{i+1}, p_{i+1}) = \Psi_{H,T}(X_i, p_i)$.
            \label{hamiltonian1}
            \State Set $(q'_{i+1}, p'_{i+1}) = \Psi_{H,T}(Y_i, p'_i)$.
            \State Sample $b_i \sim \unif([0,1])$.
            \Comment{Shared $b_i$}
            \State Set $X_{i + 1}$ = \textsc{MHAdj}($X_i$, $q_{i+1}$, $p_i$, $p_{i+1}$, $b_i$, $H$)
            \State Set $Y_{i + 1}$ = \textsc{MHAdj}($Y_i$, $q'_{i+1}$, $p'_i$, $p'_{i+1}$, $b_i$, $H$)
            \label{hamiltonian2}
        \EndFor
    \EndProcedure
\end{algorithmic}
Considering either the $X_i$'s or $Y_i$'s in isolation, this procedure
is identical to the basic HMC algorithm, so if $X_0$ and $Y_0$ are
drawn from the same initial distribution then marginally the
distributions of $X_{1:n}$ and $Y_{1:n}$ are the same. However, the
chains are not independent, since they share the random $p_i$'s and $b_i$'s.
In fact, under certain reasonable conditions\footnote{
For example, if the gradient of $U$ is sufficiently smooth and $U$ is strongly convex outside of some Euclidean ball \citep{bou-rabee}.
HMC chains will also couple if $U$ is strongly convex ``on average'' \citep{mangoubi}; presumably there are many ways of formulating sufficient conditions for coupling. That said, we cannot expect the chains to couple quickly if HMC does not mix quickly, for example if $U$ is highly multimodal.}
on $U$ it can be shown that the dynamics of the $X_i$ and the $Y_i$ approach each other.
More precisely, under these conditions each step in the Markov Chain is contractive in the sense that
\begin{equation*}
    \bbE[||X_{n+1}-Y_{n+1}||] < c\bbE[||X_n-Y_n||]
\end{equation*}
for some constant $c<1$ depending on $U$ and $T$.

Now suppose that $U$ is symmetric about some (possibly unknown) vector $\mu$, so that
we have $U(x)=U(2\mu-x)$, and we replace line~\ref{momentum} with
\begin{center}
\begin{algorithmic}[0]
\State Set $p'_i = -p_i$
\end{algorithmic}
\end{center}
We call the new algorithm HMC-ANTITHETIC.
With HMC-ANTITHETIC, the sequence $(X_1,2\mu-Y_1), (X_2,2\mu-Y_2), \ldots$ is a coupled pair of HMC chains for the distribution given by $U(x)$.
After all, it is the same as the original coupled algorithm apart from $Y_i, q'_i$ and $p_i'$ being reflected.
So both $X_i$ and $Y_i$ converge to the correct marginal distribution $\calZ^{-1}\exp(-U(x))$.
More interestingly, if $X$ and $2\mu - Y$ couple we now have that
\begin{equation*}
\bbE[||X_{n+1}+Y_{n+1}-2\mu||] < c\bbE[||X_n+Y_n-2\mu||].
\end{equation*}
This implies that for large enough $n$
\begin{equation}
\begin{split}
X_n + Y_n &= 2\mu + O(c^n)
\\ X_n - \mu &= -(Y_n - \mu) + O(c^n),
\end{split}
\end{equation}
so we expect the $X_i$ and $Y_i$ to become perfectly anti-correlated
exponentially fast.
This means that we can expect $\frac{1}{2n}\sum_i f(X_i)+f(Y_i)$ to provide very good estimators for $\bbE[f(X)]$,
at least if $f$ is monotonic \citep{craiu}.

For many target distributions of interest, $U(x) = U(2\mu-x)$ holds only approximately.
Empirically, approximate symmetry generates approximate anti-coupling, which is
nonetheless often sufficient to provide very good variance reduction
(see section~\ref{sec:experiments}).

\subsection{Control Variates}
The method of control variates is another classic variance-reduction
technique \citep{botev-ridder}. Define
\begin{equation}
\begin{split}
Z = f(X) - \beta^\top(g(Y) - \bbE[g(Y)]).
\end{split}
\end{equation}
By linearity of expectation, $\bbE[Z] = \bbE[f(X)]$ for any $\beta$
and $g$, so $Z$ is an unbiased estimator of $\bbE[f(X)]$. But, if
$f(X)$ and $g(Y)$ are correlated and $\beta$ is chosen appropriately,
the variance of $Z$ may be much lower than that of $f(X)$. In
particular,
\begin{equation}
\begin{split}
\Var(Z) &= \bbE[(f(X) - \bbE[f(X)] - \beta^\top(g(Y) - \bbE[g(Y)]))^2]
\end{split}
\end{equation}
That is, the variance of $Z$ is the mean squared error between $f(X) -
\bbE[f(X)]$ and the predictor $\beta^\top (g(Y) - \bbE[g(Y)])$. An
estimate of the optimal $\beta$ can thus be obtained by linear
regression. If this predictor is accurate, then the variance of $Z$
may be much lower than that of $f(X)$.

In the absence of an analytic value for $\bbE[g(Y)]$ we use multilevel Monte Carlo as described in \citep{botev-ridder},
plugging in an unbiased estimator $\hat g$ of $\bbE[g(Y)]$.
If the variance of $\hat g$ can be averaged away more cheaply than that
of $f(X)$, then this multilevel Monte Carlo scheme can still confer a
very significant computational benefit.

We can modify HMC-COUPLED to yield a control-variate estimator
by replacing line~\ref{hamiltonian1} with the line
\begin{center}
\begin{algorithmic}[0]
Set $(q'_{i+1}, p'_{i+1}) = \Psi_{H',T}(Y_i, p'_i)$.
\end{algorithmic}
\end{center}
and line~\ref{hamiltonian2} with the line
\begin{center}
\begin{algorithmic}[0]
Set $Y_{i + 1}$ = \textsc{MHAdj}($Y_i$, $q'_{i+1}$, $p'_i$, $p'_{i+1}$, $b_i$, $H'$)
\end{algorithmic}
\end{center}
where $H'$ is a different Hamiltonian that approximates $H$:
\begin{equation}
\begin{split}
H'(q, p) \triangleq \frac{1}{2}||p||^2 - \log Q(q)
\end{split}
\end{equation}
for a density $Q$ chosen so that $Q\approx P$ and so that
$\bbE_Q[Y]$ is tractable to compute or cheap to estimate. Finally, we
define the variance-reduced chain
\begin{equation}
\begin{split}
\label{eq:cvz}
Z_{i,j} = f_j(X_i) - \beta_j^\top (f(Y_i) - \hat \bbE_Q[f(Y)]),
\end{split}
\end{equation}
where $f$ is a vector-valued function whose expectation we are interested
in, and $\hat \bbE_Q[f(Y)]$ is an unbiased estimator of $\bbE_Q[f(Y)]$.
We call this new algorithm HMC-CONTROL.

In English, the idea is to drive a pair of HMC chains with the same
randomness but with different stationary distributions $P$ and $Q$ and
use the chain targeting $Q$ as a control variate for the chain
targeting $P$. The more closely $Q$ approximates $P$, the more
strongly the two chains will couple, the more predictable $X$ will
be given $Y$, and the lower the variance of $Z$ will become.

\subsection{Combining Antithetic and Control Variates}

HMC-ANTITHETIC and HMC-CONTROL rely on different coupling strategies,
so in principle we can get further variance reduction by combining the
two algorithms.  Specifically, we can augment the $X$ and $Y$ chains
in HMC-CONTROL with antithetic chains $X^-$ and $Y^-$, and combine
$X^-$ and $Y^-$ to get an antithetic variance-reduced chain $Z^-$ as
in \Cref{eq:cvz}. If we choose the approximating distribution $Q$ to
be symmetric about a known mean $\mu$ (for example, if $Q$ is a
multivariate Gaussian) then a further optimization is to directly set
$Y_i^- = 2\mu - Y_i$. We denote the resultant algorithm HMC-CVA, and provide a
listing of it in \Cref{sec:hmc_cva}.
\Cref{fig:traces} shows that HMC-CVA does induce the variance-reduced chains
$Z^+$ and $Z^-$ to anti-couple, leading to a further variance reduction.

\subsection{Preconditioning $U$}
\label{sec:preconditioning}

HMC-CVA works best when $P$ is approximately symmetric, and can be
approximated by a tractable symmetric distribution $Q$ with known
mean.  Clearly not all target distributions satisfy this condition.
Following \citet{parno2014transport,hoffman2019neutra}, we can
partially resolve this issue by applying a change of variables $\tilde
x \triangleq m^{-1}(x)$ to $P$ such that $P(\tilde
x)=P(x)|\frac{\partial m}{\partial \tilde x}| \approx
\mathcal{N}(\tilde x; 0, I)\triangleq Q(\tilde x)$, and running our
samplers in $\tilde x$ space instead of $x$ space. This also increases
HMC's mixing speed.

The method is agnostic to how we obtain the transport map $m$; in our
experiments we use variational inference with reparameterization
gradients \citep[e.g.; ][]{kucukelbirADVI} to minimize the
Kullback-Leibler (KL) divergence from $Q(x)$ to $P(x)$ (or
equivalently from $\mathcal{N}(0, I)$ to  $P(\tilde x)$).

\section{Related Work}
\subsection{Markov Chain coupling}
Coupling has long been used as a theoretical tool to study and prove convergence behavior of Markov chains and therefore Markov Chain Monte Carlo (MCMC) algorithms \citep{lindvall1992lectures}.
For example \cite{bou-rabee} use coupling to show that the steps in the HMC algorithm are contractive.
Coupling has also been used to derive practical algorithms.
\citet{glynn2014exact}, \citet{jacob2017unbiased}, and \citet{heng2019unbiased} use Markov chain coupling to give unbiased estimators.

\subsection{Markov Chain Monte Carlo and Antithetic Variates}
In early work in this area, \cite{hastings1970monte} uses an
alternating sign-flip method, similar to antithetic variates, to
reduce the variance of samples from generated with a single Markov
Chain targeting a symmetric toy distribution.
\citet{green1992metropolis} further develop of this method, still
using a single Markov chain.

\citet{frigessi2000antithetic} propose an antithetic Gibbs sampling
scheme very similar in spirit to antithetic HMC, in which each
one-dimensional complete conditional distribution is sampled by CDF
inversion driven by antithetic uniform noise.  Antithetic HMC has
three main advantages over antithetic Gibbs: first, HMC often mixes
much more quickly than Gibbs sampling (so there is less variance to
eliminate in the first place); second, HMC can be applied to models
with intractable complete-conditional distributions
(\citet{frigessi2000antithetic} found that switching to a
Metropolis-within-Gibbs scheme to handle intractable complete
conditionals reduced the variance-reduction factor to 2, which is no
better than what could be obtained by running an independent chain);
finally, HMC is more amenable than Gibbs to preconditioning schemes
\citep{parno2014transport,hoffman2019neutra}, which accelerate both
mixing and coupling.

\subsection{Coupling and Control Variates}
In previous work the method of control variates has been proposed as a
method to reduce variance in MCMC methods.  However, control variates
require the use of an approximating distribution whose properties can
be computed either analytically, or much more efficiently than the
target distribution.  \citet{dellaportas2012control} note that finding
such distributions has posed a significant challenge.  That paper
provides a general purpose method for a single Markov chain that can
be used whenever we have an explicit computation of
$\bbE[g(X_{n+1})|X_n=x]$ for a Markov chain $\{X_i\}$ and for certain
linear functions $g$.  \citet{andradottir1993variance} use a smoothing
method to improve the efficiency of Monte Carlo simulations, and use
approximations to smoothing to derive control variates.
\citet{baker2019control} use control variates to reduce the variance
arising from stochastic gradients in Langevin dynamics.

\citet{goodman2009coupling} use a coupled pair of Markov chains, both
of which have marginal distributions equal to the target distribution.
The second of these chains is then used to provide proposals for a
Metropolis-Hastings algorithm that generates samples from an
approximation to the target distribution.  It is this approximate
distribution which serves as their control variate.

Our control-variate HMC scheme can be seen as an adaptation of the
method of \citet{pinto2001improving}, who only considered coupling
schemes based on Gibbs sampling and random-walk Metropolis, neither of
which typically mixes quickly in high dimensions
\citep{neal2011mcmc}. The Gibbs scheme was reported to couple
effectively, but is limited to problems with tractable complete
conditionals, while the random-walk Metropolis scheme required a small
step size (which dramatically slows mixing) to suppress the rejection
rate and ensure strong coupling (an issue we will see in section
\ref{sec:experiments}). By contrast, HMC is applicable to any
continuously differentiable target distribution, and its rejection
rate can be kept low without sacrificing mixing speed.

\section{Experiments}
\label{sec:experiments}

We test the proposed variance reduction schemes on three posterior distributions.

\subsection{Target Distributions}

\begin{figure*}[!tb]
  \centering
    \includegraphics[scale=0.7]{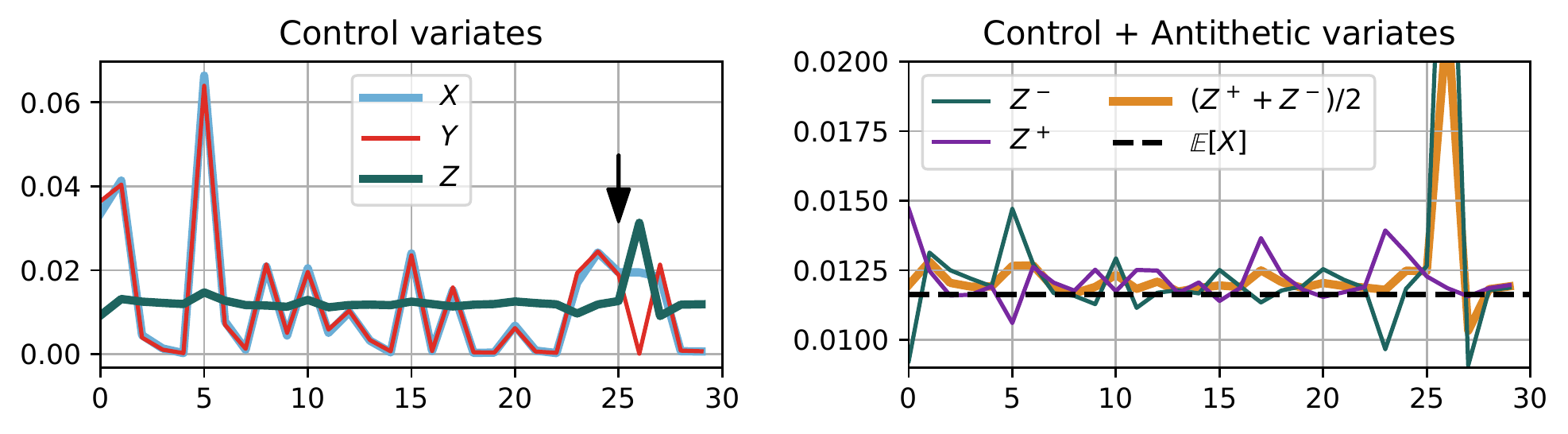}
\caption{
    On the left, a trace of 30 steps of the HMC chain $X_i$ with its coupled
    chain $Y_i$. Both chains are estimating the posterior variance of a single
    dimension for the logistic regression posterior. By using
    $Y_i$ as a control variate to $X_i$, we obtain a variance-reduced chain
    $Z$. The coupling is strong, and is only interrupted when the two chains
    make opposite decisions in the Metropolis-Hastings correction (arrow). On
    the right, we can average $Z$ (labeled $Z^+$ for emphasis) and its antithetically coupled chain
    $Z^-$ to further reduce the variance of the estimator.
}
\label{fig:traces}
\end{figure*}

\paragraph{Logistic regression:}

We first consider a simple Bayesian logistic regression task, with the model
defined as follows:
\begin{equation}
\label{eq:logistic}
\begin{split}
w_d &\sim \calN(0, 1);\quad
y_n \sim \Bern(\sigma(x_n^\top w))
\end{split}
\end{equation}
where $w$ are the covariate weights, $\Bern$ is the Bernoulli distribution and
$\sigma$ is the sigmoid function. We precondition the sampler using a full-rank affine transformation.

We apply the model to the German credit dataset \citep{dua2019uci}. This
dataset has 1000 datapoints and 24 covariates (we add a constant covariate for
the bias term, bringing the total to 25) and a binary label. We standardize the
features to have zero mean and unit variance.

\paragraph{Sparse logistic regression:}

To illustrate how variance reduction works with non-linear preconditioners and
complicated posteriors we consider a hierarchical logistic regression
model with a sparse prior:
\begin{equation}
\begin{split}
\tau &\sim \Gam(\alpha=0.5, \beta=0.5) \\
\lambda_d &\sim \Gam(\alpha=0.5, \beta=0.5) \\
w_d &\sim \calN(0, 1);\quad 
y_n \sim \Bern(\sigma(x_n^\top (\tau w \circ \lambda)))
\end{split}
\end{equation}
where $\Gam$ is the Gamma distribution, $\tau$ is the overall scale,
$\lambda$ are the per-component scales,
$w$ are the non-centered covariate weights, and $w \circ \lambda$ denotes the
elementwise product of $w$ and $\lambda$. The sparse gamma prior on $\tau$ and
$\lambda$ imposes a soft-sparsity prior on the overall weights. This
parameterization uses D=51 dimensions. Following \citet{hoffman2019neutra},
we precondition the sampler using a neural-transport (NeuTra)
stack of 3 inverse autoregressive flows (IAFs) \citep{kingma2016iaf} with 2
hidden layers per flow, each with 51 units, and the ELU nonlinearity
\citep{clevert2015fast}.

\paragraph{Item Response Theory}

As a high-dimensional test, we consider the 1PL item-response theory model.
\begin{equation}
\begin{split}
    \delta \sim \calN(0.75, &1);\quad
    \alpha_i \sim \calN(0, 1);\quad
    \beta_j \sim \calN(0, 1)
\\
    y_{ij} &\sim \Bern(\sigma(\alpha_i - \beta_j + \delta))\,\vert i,j \in \mathcal{T}
\end{split}
\end{equation}
where $\delta$ is the mean student ability, $\alpha_i$ is ability of student
$i$, $\beta_j$ is the difficulty of question $j$, $y_{ij}$ is whether a question
$j$ was answered correctly by student $i$ and $\mathcal{T}$ is the set of
student-question pairs in the dataset.  We precondition the sampler using a
full-rank affine transformation.

We apply the model to the benchmark dataset from the Stan example
repository\footnote{\url{https://github.com/stan-dev/example-models}} which has
400 students, 100 unique questions and a total of 30105 questions answered. In
total, the posterior distribution of this model has D=501 dimensions.

\subsection{Sampling procedure}

As described in \Cref{sec:preconditioning},
we train a transport map $m$ by variational inference with
reparameterization gradients to minimize $\mathrm{KL}(Q||P)$,
where $Q(x)=\mathcal{N}(m^{-1}(x); 0, I)|\frac{\partial m^{-1}}{\partial x}|$
\citep{kucukelbirADVI}.
The resultant distribution is used as the initial state
for the HMC sampler. To compute accurate estimates of performance, we run a few
thousands of chains for 1000 steps, discarding the first 500 to let the chains
reach their stationary distribution.

We tune the hyper-parameters of the HMC sampler to maximize the effective
sample size (ESS) normalized by the number of gradient evaluations of $U$,
while making sure the chain remains stationary. ESS is a measure of the
variance of MCMC estimators, defined as $ESS = N
\Var[\hat{\bbE}_{MC}[f(X)]] / \Var[\hat{\bbE}_{MCMC}[f(X)]]$.

\begin{figure*}[!tb]
  \centering
    \includegraphics[scale=0.75]{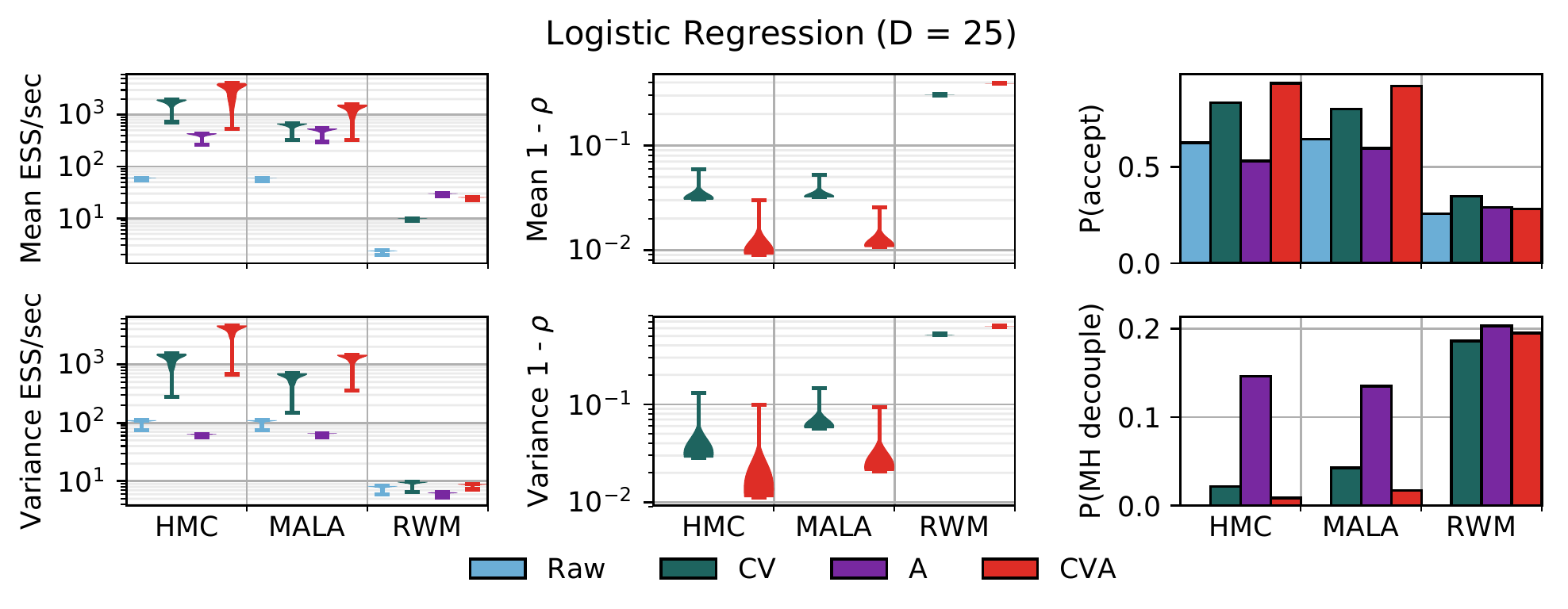}
    \includegraphics[scale=0.75]{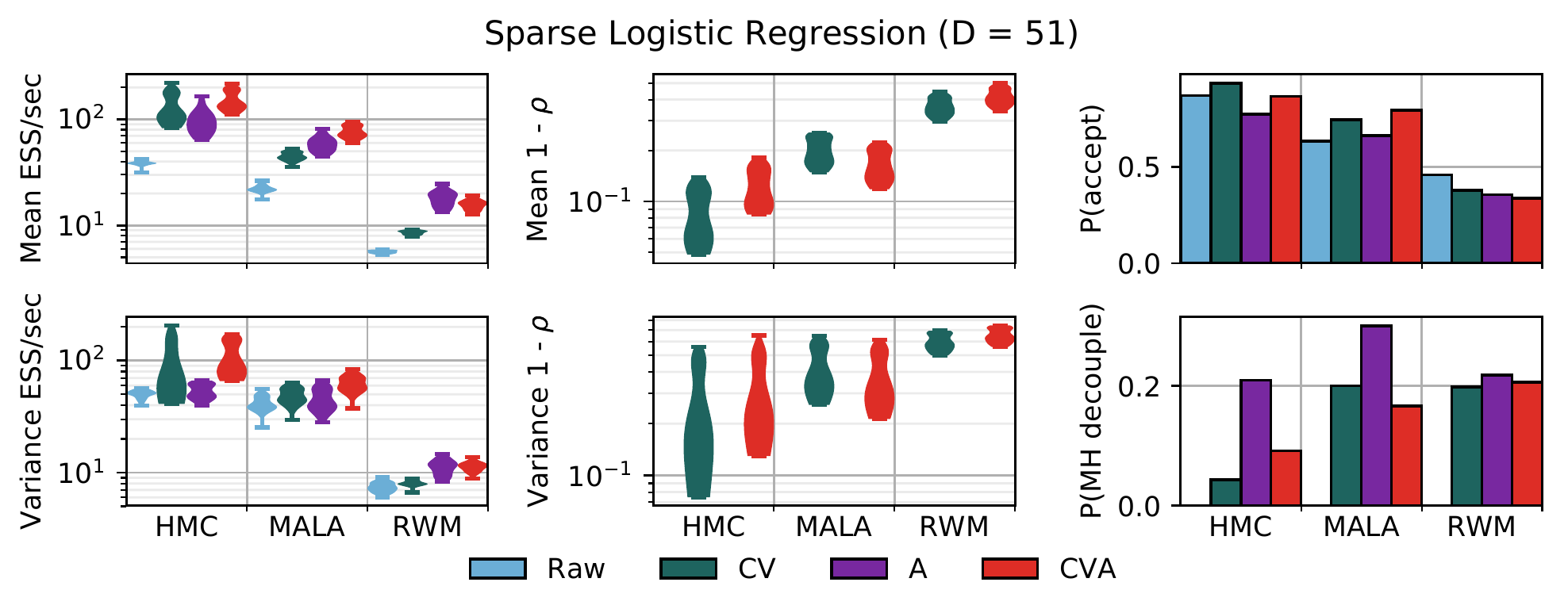}
    \includegraphics[scale=0.75]{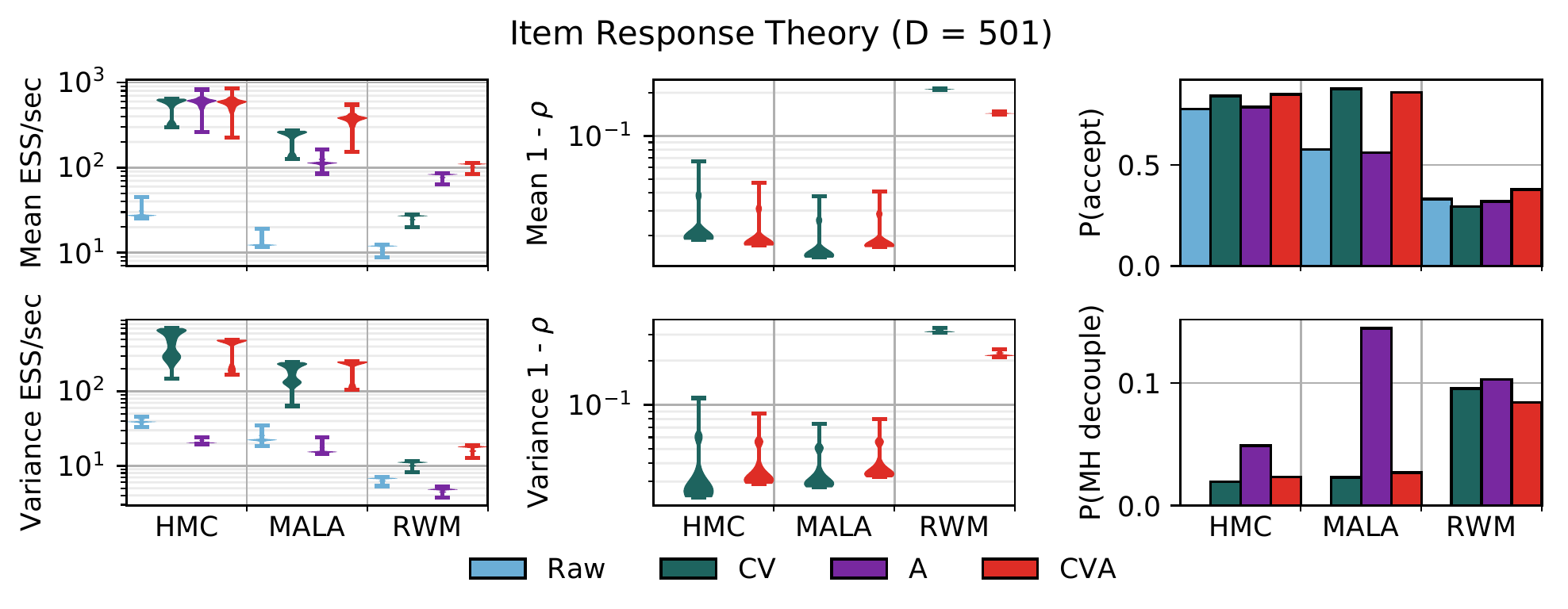}
\caption{
    Performance of the various variance reductions schemes when estimating the
    posterior mean and variance for the two target distributions. First column
    shows violin plots (aggregated across the dimensions of $X$) of ESS/sec
    (log scale), each algorithm tuned as described in the main text. Second
    column shows the correlation achieved at the optimal set of hyperparameters
    (log scale).
    Top panel of third column shows the acceptance probabilities of the MCMC
    chains, while the bottom panel shows the probability of two chains
    decoupling due to the Metropolis-Hastings adjustment step. For all
    quantities, we compute the mean across 10 independent runs.
}
\label{fig:ess}
\end{figure*}

We measure chain stationarity using the potential scale reduction metric
\citep{gelman1992}. The ESS is estimated using the autocorrelation method with
the Geyer initial monotone sequence criterion to determine maximum lag
\citep{geyer2011introduction, vehtari2019rank}.

We estimate and use $\beta$ for all the chains. This introduces some bias in
the estimator, although the estimator remains consistent
\citep{pinto2001improving}. All experiments were run on a Tesla P100 GPU. The
source code is available at
\url{https://github.com/google-research/google-research/tree/master/hmc_swindles}.

\subsection{Results}

\begin{figure*}[!tb]
  \centering
    \includegraphics[scale=0.75]{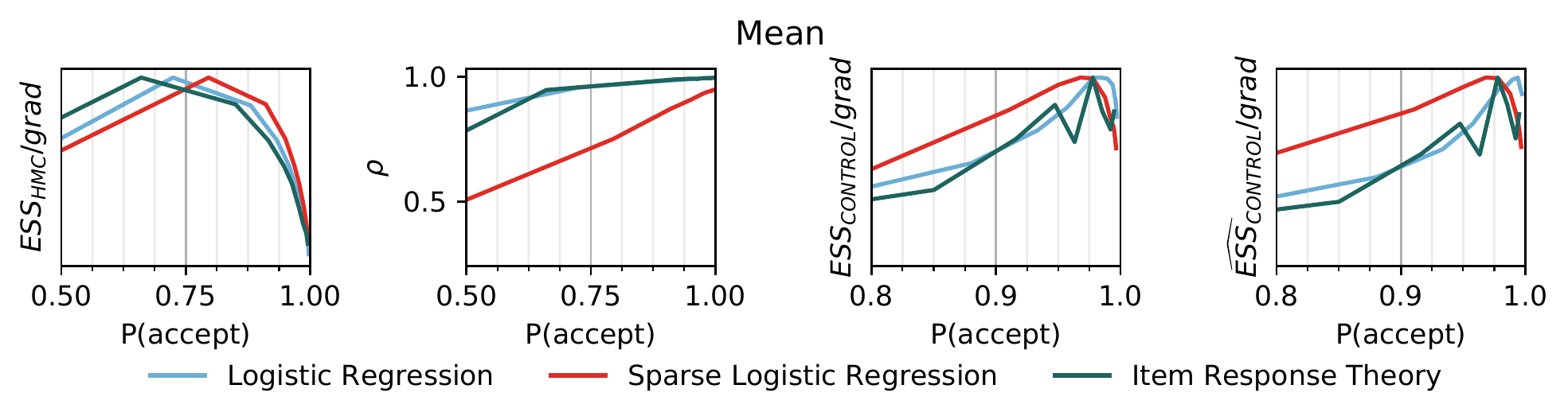}
    \includegraphics[scale=0.75]{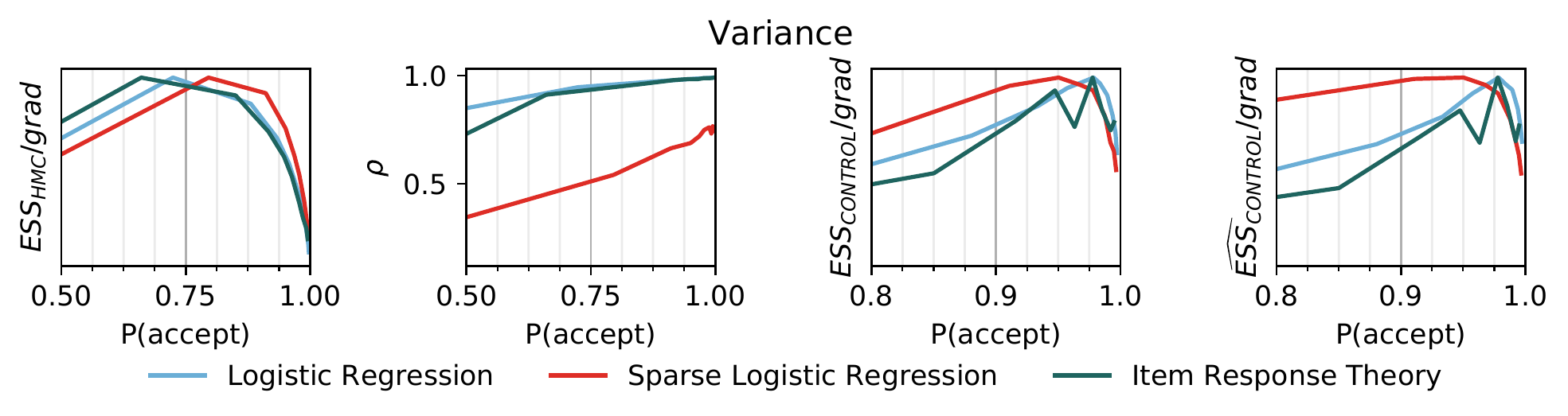}
\caption{
    Effect on ESS/grad and $\rho$ when varying the number of leapfrog steps
    while keeping HMC trajectory length constant for different functions (rows)
    and probability distributions (different lines). For raw HMC, the optimal
    ESS/grad is obtained by choosing an acceptance probability in the 0.6-0.9
    regime (first column). For HMC-CONTROL decreasing the step size raises
    $\rho$ (second column) and ESS/grad (third column), up to a point where the
    original chain efficiency drops below what variance reduction can
    compensate for, at an acceptance probability near 0.95. As a tuning
    heuristic, we can predict the location of the optimal acceptance rate by
    leaning on the theoretical bounds on ESS/grad and empirical $\rho$ estimate
    (final column).
}
\label{fig:bounds}
\end{figure*}

\begin{figure}[!tb]
  \centering
    \includegraphics[scale=0.75]{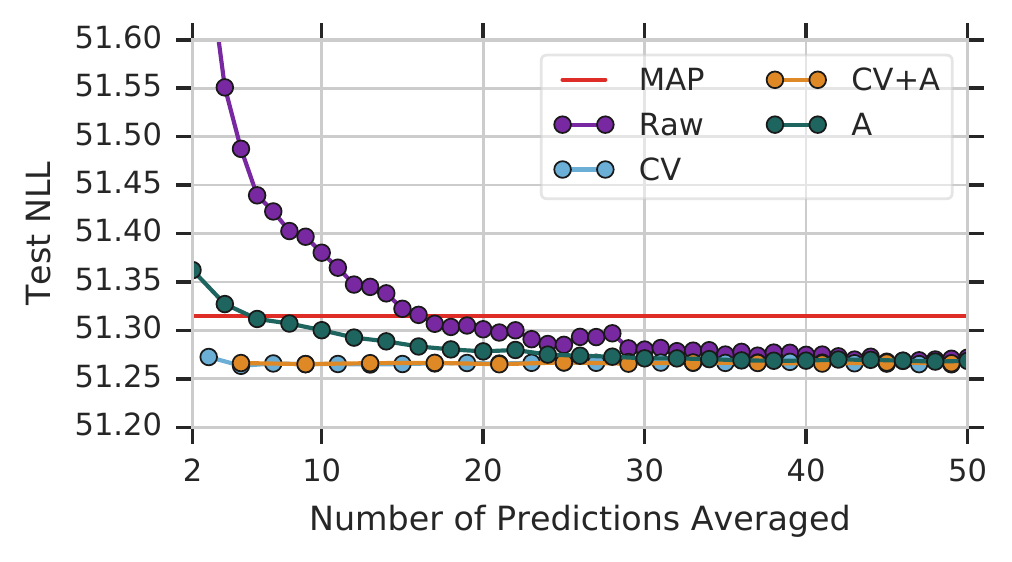}
\caption{
    Median held-out test negative log-likelihood for logistic regression, lower
    is better. The x-axis measures the number of averaged models, including
    the cost of evaluating any antithetic samples, control variates, and expectations
    under $Q$ (assuming the latter is available in closed form). The MAP
    estimate of the parameters provides a variance-free estimate, but it is
    suboptimal compared to a fully Bayesian model average. Without variance reduction,
    the raw HMC estimator requires averaging many independent samples to reach parity with
    the MAP estimator. The HMC-CONTROL and HMC-CVA estimators make accurate predictions
    with very few function evaluations; HMC-ANTITHETIC is less efficient, but still
    much faster than independent samples.
}
\label{fig:test_nll}
\end{figure}

For simple $P$'s like the logistic-regression posterior distribution, we
observe that $X$ and its control variate $Y$ couple very well leading to
dramatic variance reduction in the $Z$ chain (\Cref{fig:traces}, left
panel). The two chains only decouple when the Metropolis-Hastings decision is
different between the two chains. Note that, since the uniform $b_i$'s are
shared, not every rejection leads to decoupling.

Adding an antithetic control-variate chain $Z^-$ further reduces the
variance of the estimator (\Cref{fig:traces}, right panel). As hoped,
the antithetic variance-reduced chains $Z^-$ are clearly anti-correlated.

Moving beyond qualitative results, we examine ESS/sec values
for various algorithms and test problems, evaluated on
estimators of the means and variances of the posteriors. For each
distribution, we run 10 independent evaluation runs, average across the runs
and then report the ESS/sec for each dimension, summarized by a violin plot
(\Cref{fig:ess}, first column). 
We
normalize ESS by wall-clock running time to control for running
additional chains (or taking different numbers of leapfrog steps for HMC).
Running antithetic chains is more expensive than the control-variate
chains, as the latter use a cheaper-to-evaluate Hamiltonian (e.g. a
Gaussian vs the full model joint distribution) We
compare our proposed schemes to the analogous schemes described in earlier work
\citep{frigessi2000antithetic,pinto2001improving}, applied to
Metropolis-adjusted Langevin algorithm (MALA)\citep{roberts1998optimal} and
Random-Walk Metropolis (RWM) \citep{metropolis1953equation}.

Overall, the variance reduction schemes improve upon the non variance-reduced
algorithms, with the control variates combined with antithetic variates
typically outperforming the two techniques individually. Our HMC-based and
MALA schemes outperform RWM by over an order of magnitude. This can in part be
explained by the high correlations attainable by HMC and MALA (\Cref{fig:ess},
second column), in part because unlike RWM, they can be efficient at high
acceptance rates (\Cref{fig:ess}, third column, top panel). A big source of
de-correlation are the Metropolis-Hastings adjustment disagreements, which
happen much less frequently for HMC and MALA than in RWM (\Cref{fig:ess},
third column, top panel). HMC-based schemes tend to be a few times more
efficient than MALA.

A weakness of the antithetic schemes is that they provide less benefit
when the the estimated function is symmetric about the mean. In
particular, $(\sigma(x^\top w) - \bbE_P[\sigma(x^\top w)])^2$ is close to being
an even function in $w - \bbE_P[w]$. This causes the antithetic schemes to be
worse than regular MCMC when estimating the variance for the logistic
regression and item response theory posteriors, but in case of the asymmetric
distribution for sparse logistic regression this is less pronounced. Despite
this, in most cases combining control variates and antithetic variates yields
some efficiency gains even when estimating variance.

The sparse logistic regression is a significantly more difficult problem, and
correspondingly we do not see as much benefit from the variance-reduction
techniques. Even with the powerful IAF approximating distribution, the
correlation between the coupled chains is not very high, but we still get
a several-fold reduction in variance.

The ESS of the variance reduced chain can be related to the ESS of the
original chain as
\begin{align}
    \label{eq:ess_vr}
    ESS_{CONTROL} &= ESS_{HMC} / (1 - \rho^2) \\
    ESS_{ANTITHETIC} &= 2 ESS_{HMC} / (1 + \rho),
\end{align}
for HMC-CONTROL and HMC-ANTITHETIC respectively, where $\rho$ is the
correlation coefficient between the corresponding pairs of chains. This suggests
that efficiency can be improved by increasing the correlation between the
paired chains. This can be done by reducing the step size, but
doing so also reduces the efficiency of the original chain. In practice there
is exists some optimal step size smaller than is optimal for a single HMC
chain, but larger than would maximize the correlation, the exact numerical
value of which depends on the details of $f(\cdot)$ and the approximating
distributions. We examine this tradeoff empirically by finding an optimal
trajectory length (same as in \Cref{fig:ess}), and then varying the number of
leapfrog steps and step size to hold trajectory length constant.
This study confirms the
basic intuition above (\Cref{fig:bounds}, first three columns). In practice, we
recommend targeting an acceptance rate in the vicinity of 0.95 for
distributions similar to ones examined here.

When explicit tuning is desired, we propose a heuristic based on replacing the
$ESS_{orig}$ in \Cref{eq:ess_vr} with an upper bound on the ESS from
\citep{beskos2010optimal}:
\begin{equation}
    \label{eq:ess_bound}
    ESS_{HMC} / grad \leq C P(accept) (\Phi(1 - P(accept)/ 2))^{0.5},
\end{equation}
where $C$ depends on the distribution, $f(\cdot)$ and the HMC trajectory length
and then doing a number of pilot runs to estimate $\rho$. Unlike ESS, $\rho$
can be readily estimated using comparatively few samples, and when plugged
into \Cref{eq:ess_vr} together with \Cref{eq:ess_bound} yields a quantity
$\widehat{ESS}_{CONTROL} / grad$ captures the maximum efficiency as a function of
acceptance probability (\Cref{fig:bounds}, last column).

A practical application of the variance-reduction schemes is to reduce the cost of accurate
predictions on new data. While Bayesian model averaging can provide good
accuracy, if it is implemented via MC sampling each test-time example may
have to be evaluated many times to reduce the variance of the ensemble prediction.
With HMC swindles, we expect this to be less of a problem. To test
this use case, we split the German credit dataset randomly into train and test
sets using a 90\%/10\% split, and then compared the test set negative
log-likelihood between the various estimators (\Cref{fig:test_nll}). Raw HMC
requires averaging many predictions to reach an acceptable level of
variance, while CV-HMC immediately produces accurate predictions.

\section{Discussion}

We have presented two variance reduction schemes for HMC based on approximate
coupling. Despite the added computation, combining both
schemes results in improved effective sample size rates. An alternate way of
interpreting this result is that when running multiple chains in parallel, converting
some of those chains to control variates and antithetic variates is a more
efficient way of using a fixed compute budget.

Our experiments have not supported broad effectiveness of the antithetic scheme
when used in isolation, in part because it relies on the hard-to-control
symmery of the target density. Despite this, it is very easy to implement and
there seems little reason not to use it. The control variate scheme is
generally applicable, but relies on the practicitioner being able to construct
efficient and accurate approximate distributions. This is an impediment to
broad use, but as more work is done to make this process automatic through the
use of powerful invertible neural networks we hope this becomes less of an
issue in the future. Overall, these techniques promise to yield estimates of
posterior expectations with extremely low bias and variance, bringing the dream
of near-exact posterior inference closer to reality.

\bibliography{main}

\newpage
\begin{appendices}
\crefalias{section}{appsection}

\section{HMC-CVA Algorithm}
\label{sec:hmc_cva}
We denote the original chain $X$ and its corresponding control variate $Y$ as
$X^+$ and $Y^+$ for emphasis.
\begin{algorithmic}[1]
    \Procedure{HMC-CVA}{$X^+_0, X^-_0, Y^+_0, n, T$}
        \For{$i=0$ to $n-1$}
            \State Sample $p^+_i \sim \calN(0, I_D)$.
            \State Set $p^-_i=-p^+_i$.
            \State Set $p'_i=p^+_i$.
            \Comment{Shared $p_i$}
            \State Set $(q^+_{i+1}, p^+_{i+1}) = \Psi_{H,T}(X^+_i, p^+_i)$.
            \State Set $(q^-_{i+1}, p^-_{i+1}) = \Psi_{H,T}(X^-_i, p^-_i)$.
            \State Set $(q'_{i+1}, p'_{i+1}) = \Psi_{H',T}(Y^+_i, p'_i)$.
            \State Sample $b_i \sim \unif([0,1])$.
            \Comment{Shared $b_i$}
            \State Set $X^+_{i + 1}$ = \textsc{MHAdj}($X^+_i$, $q^+_{i+1}$, $p^+_i$, $p^+_{i+1}$, $b_i$, $H$).
            \State Set $X^-_{i + 1}$ = \textsc{MHAdj}($X^-_i$, $q^-_{i+1}$, $p^-_i$, $p^-_{i+1}$, $b_i$, $H$).
            \State Set $Y^+_{i + 1}$ = \textsc{MHAdj}($Y^+_i$, $q'_{i+1}$, $p'_i$, $p'_{i+1}$, $b_i$, $H'$).
            \State Set $Y^-_{i + 1} = 2\mu - Y^+_{i+1}$.
        \EndFor
        \State Compute unbiased estimate $\hat\bbE_Q[f(Y)]$.
        \State Estimate optimal $\beta$ by linear regression.
        \For{$i=1$ to $n$}
        \State Set $Z^+_{i,j} = f_j(X^+_i) - \beta_j^\top (f(Y^+_i) - \hat\bbE_Q[f(Y)])$.
        \State Set $Z^-_{i,j} = f_j(X^-_i) - \beta_j^\top (f(Y^-_i) - \hat\bbE_Q[f(Y)])$.
        \State Set $Z_i = \frac{1}{2}(Z^+_i + Z^-_i)$.
        \EndFor
    \EndProcedure
\end{algorithmic}

\end{appendices}

\end{document}